\documentclass[12pt,a4paper]{article}
\pdfoutput=1

\usepackage{jcappub}
\usepackage{soul}
\allowdisplaybreaks

\usepackage[applemac]{inputenc}

\usepackage[margin=1in]{geometry}
\usepackage{graphicx}
\usepackage{amsfonts}
\usepackage{subfigure}
\usepackage{float}
\usepackage{hyperref}
\usepackage[dvipsnames]{xcolor}
\usepackage{amsmath}
\usepackage[normalem]{ulem}
\usepackage{empheq}
\usepackage{amssymb}
\usepackage{tikz}
\usepackage{tabularx}

\usepackage{booktabs}

\def\lsim{~\rlap{$<$}{\lower 1.0ex\hbox{$\sim$}}}
\def\bsim{~\rlap{$>$}{\lower 1.0ex\hbox{$\sim$}}}

\def\ln{{\rm ln}}

\def\be{\begin{equation}}
\def\ee{\end{equation}}
\def\bea{\begin{eqnarray}}
\def\eea{\end{eqnarray}}
\def\ba{\begin{align}}
\def\ea{\end{align}}
\def\bi{\begin{itemize}}
\def\ei{\end{itemize}}

\newcommand{\bn}{{\mathbf n}}

\def\bx{{\bf{x}}}
\def\bk{{\bf{k}}}

\def\bv{{\bf{v}}}

\def\ba{{\vec{g}}}

\newcommand{\HH}{\mathcal{H} }

\newcommand{\ndv}{v_{||}}
\newcommand{\dotndv}{\dot{v}_{||}}

\newcommand{\ndvthree}{{v_{||}^{(3)}}}

\title{The relativistic galaxy number counts in the weak field approximation}

\author[a,b,c]{Enea~Di~Dio}
\author[d,e]{Florian~Beutler}

\affiliation[a]{Center for Theoretical Astrophysics and Cosmology, Institute for Computational Science, University of Zurich, Winterthurerstrasse 190, CH-8057 Zurich, Switzerland}
\affiliation[b]{Physics Division, Lawrence Berkeley National Laboratory, Cyclotron Rd, Berkeley, CA 94720}
\affiliation[c]{Berkeley Center for Cosmological Physics and Department of Physics, University of California, Berkeley, CA 94720}
\affiliation[d]{Institute for Astronomy, University of Edinburgh, Royal Observatory, Blackford Hill, Edinburgh EH9 3HJ, UK
}
\affiliation[e]{Institute of Cosmology \& Gravitation, University of Portsmouth,
Dennis Sciama Building, Burnaby Road, Portsmouth PO1 3FX, UK}

\abstract{
We present a novel approach to compute systematically  the relativistic projection effects at any order in perturbation theory within the weak field approximation. In this derivation the galaxy number counts is written completely in terms of the redshift perturbation. The relativistic effects break the symmetry along the line-of-sight and they source, contrarily to the standard perturbation theory, the odd multipoles of the matter power spectrum or 2-point correlation function, providing a unique signature for their detection in Large Scale Structure surveys. We show that our approach agrees with previous derivations (up to third order) of relativistic effects and, for the first time, we derive a model for the transverse Doppler effect. Moreover, we show that in the Newtonian limit this approach is consistent with standard perturbation theory at any order.
}

\emailAdd{enea.didio@uzh.ch}
\emailAdd{florian.beutler@port.ac.uk}

\begin{document}
\maketitle

%%%%%%%%%%%%%%%%%%%%%%%%%%%%%%%%%%%%%%%%%%%%%%%%%
\section{Introduction}

The so-called relativistic projection effects arise form the change of coordinates required to describe galaxy redshift survey observables directly in terms of observed quantities, namely the redshift $z$ and the line-of-sight direction $\bn$. In doing this change of coordinates we have to carefully consider that both these quantities are perturbed with respect to the background cosmology. While current derivations and expressions of relativistic corrections are long and tedious (see for instance Refs.~\cite{Yoo:2009,Yoo:2010,Bonvin:2011bg,Challinor:2011bk,Hall:2012wd,Yoo:2014sfa,Bertacca:2014dra,Bertacca:2014wga,DiDio:2014lka,Irsic:2015nla,DiDio:2016ykq,DiDio:2018zmk,Clarkson:2018dwn,Maartens:2019yhx}), one should be able to recover them directly from the shift in redshift $\delta z $ and in the line-of-sight $\delta \bn$, at least the leading non-vanishing order in the weak field approximation.

Moreover, this approach provides a clear interpretation in terms of numerical N-body simulations. Indeed, it is worth reminding that relativistic projection effects are purely geometrical and independent of the theory of gravity. Because of that, these effects can be correctly computed using geodesic light tracing in a Newtonian N-body simulation, see e.g.~Ref.~\cite{Breton:2018wzk}.

Intuitively we can associate the redshift $\delta z$ and the line-of-sight $\delta \bn$ perturbations to radial and transverse modes, respectively. The shift in transverse position $\delta \bn$ is generated by the lensing effect of foreground structures and is a well-known effect studied and detected for several different observables: CMB (see Ref.~\cite{Lewis:2006fu} for an extensive review), magnification and shear galaxies (see reviews~\cite{Bartelmann:1999yn,Schneider:2005ka,Kilbinger:2014cea}), CIB~\cite{Schaan:2018yeh}, $21$cm intensity mapping~\cite{Mandel:2005xh} and matter power spectrum~\cite{Hui:2007cu,LoVerde:2007ke,Hui:2007tm,Dodelson:2008qc,Montanari:2015rga,DiDio:2016kyh}.
In this work we focus on the effects generated by the redshift perturbation $\delta z$, which is sourced by peculiar motion of galaxies and geometrical perturbations in the form of gravitational redshift and integrated Sachs-Wolf effect (ISW). The relativistic projection effects generated by the redshift perturbation peak along the radial direction, e.g. $\mu \equiv - \bn \cdot \hat\bk = \pm1$, where $\bk$ is the Fourier wave vector associated with the pair separation and $\bn$ is the line-of-sight unit vector pointing from the source to the observer. The largest effect induced by the redshift perturbation is the well-known Kaiser redshift space distortion~\cite{Kaiser:1987qv}. While the latter is proportional to $\mu^2$, relativistic effects generate terms proportional to odd powers of $\mu$. Therefore, differently from standard perturbation theory, relativistic effects source the odd multipoles of the matter power spectrum or 2-point correlation function, as first pointed out in Ref.~\cite{McDonald:2009ud} and then studied for different observables~\cite{Yoo:2012,Bonvin:2013ogt,Irsic:2015nla,Gaztanaga:2015jrs,Bonvin:2015kuc,Hall:2016bmm,Bonvin:2016dze,Giusarma:2017xmh,Lepori:2017twd,Breton:2018wzk}.

In this work we develop a formalism to easily obtain the relativistic corrections which source the odd multipoles to any order in perturbation theory, omitting integrated terms whose magnitudes are expected to be negligible in terms of the odd multipole statistics, in most of the relevant configurations. 
Our work provides the necessary tools to compute the odd-multipole statistics of the power spectrum or the 2-point correlation function of the Large Scale Structures (LSS) beyond linear theory.

In an accompanying paper~\cite{Beutler:2020}, we adopt this formalism to compute the dipole of the matter power spectrum at 1-loop in perturbation theory and compare our theoretical prediction with the relativistic N-body simulation presented in Ref.~\cite{Breton:2018wzk}.

The paper is organized as follows: in Section~\ref{sec:rel_eff} we introduce the relativistic effects within the weak field approximation and then, in Section~\ref{sec:redshift}, we compute the full redshift perturbation within this approximation. In Sections~\ref{sec:density} and~\ref{sec:volume} we compute the density and the volume perturbations, respectively, sourced by the redshift perturbation $\delta z$. In Section~\ref{sec:full_counts} we derive the full galaxy number counts and we compare with previous derivations up to third order in perturbation theory. In Section~\ref{sec:single_contributions} we derive the relativistic effects induced directly by the gravitational redshift, the linear and transverse Doppler in the redshift perturbation. In Section~\ref{sec:newt_limit} we proof that in the Newtonian limit our approach fully agrees with standard perturbation theory at any order and we conclude in Section~\ref{sec:conclusions}.

\section{Relativistic effects}
\label{sec:rel_eff}

A galaxy clustering survey maps the galaxy distribution $N\left( \bn , z \right) $ as a function of line-of-sight direction $\bn$ and redshift $z$. From this we can define the galaxy number counts as
\be
\Delta \left( \bn , z \right)  = \frac{N \left( \bn , z \right) - \langle N \left( \bn , z \right) \rangle}{\langle N \left( \bn , z \right) \rangle}
\ee
where $\langle .. \rangle$ denotes the angular average at constant observed redshift $z$.

Starting from a density field $\rho \left( t, \bx \right)$ we need to perform a change of coordinates to determine the galaxy number counts $\Delta \left( \bn , z \right) $ as a function of the observed angle $\bn$ and redshift $z$.
The matter density $\rho \left( t, \bx \right)$ will transform under a change of coordinates as a density scalar field: through a density and a volume part, studied in Sections~\ref{sec:density} and~\ref{sec:volume}, respectively. 
This change of coordinates requires to solve the geodesic light propagation from the source to the observer. Therefore, the galaxy number counts $\Delta \left( \bn , z \right) $ will also be affected by the matter distribution between the source and the observer. Moreover, it will also be sensitive to the peculiar motion of the source  which impacts the redshift measurement. The leading part of the dependence on the peculiar motion of the source is captured by the redshift space distortion effect~\cite{Kaiser:1987qv}. However, several other subleading contributions are not captured in the Newtonian derivation of the redshift space distortion effect and we usually refer to them as relativistic corrections.

More precisely, we decompose the galaxy number counts in a Newtonian part $\Delta_N$ and a (leading) relativistic correction $\Delta_R$ as
\be
\Delta = \Delta_N + \Delta_R +\mathcal{ O} \left( \epsilon_\HH \right) 
\ee
where the relativistic part is suppressed as
\be
\Delta_R \sim  \epsilon_\HH^{1/2} \Delta_N \,  \quad \text{and} \quad \epsilon_\HH\equiv \left( \frac{\HH}{k} \right)^2 \ ,
\ee
where $\HH$ is the comoving Hubble parameter and $k$ the Fourier wavelength.
We remark that under this definition we do not consider lensing magnification as a relativistic effect. The impact of lensing magnification on galaxy clustering has been already extensively studied (see e.g.~Refs.~\cite{Hui:2007cu,LoVerde:2007ke,Hui:2007tm,Dodelson:2008qc,Raccanelli:2013gja,Montanari:2015rga,Alonso:2015uua,Raccanelli:2015GR,Cardona:2016qxn,Lorenz:2017iez,Villa:2017yfg}). This definition of relativistic effects indicates that corrections to the standard (Newtonian) perturbation theory could be relevant at large scales, e.g.~when these corrections are less suppressed by the parameter $\epsilon_\HH$. However, the detection of relativistic effects at the largest scales is strongly limited by cosmic variance. The parameter $\epsilon_\HH$ corresponds to the inverse Laplacian in real space. Therefore, the relativistic $\Delta_R$ and the Newtonian $\Delta_N$ parts differ by a single spatial derivative. As pointed out before, the redshift perturbation $\delta z$ is mainly induced by radial modes. Hence, as we will see in detail in the next sections, $\Delta_N$ and $\Delta _R$ differ by a radial derivative or a factor $i \mu \HH/k $ in Fourier space. 
Therefore, relativistic terms $\Delta_R$ differ also by a factor of $\mu$ with respect to the standard Newtonian terms $\Delta_N$. Since the latter are characterized only by even powers of $\mu$ (i.e.~are symmetric with respect to the angle between the line-of-sight and the pair separation) the relativistic corrections are proportional to odd powers of $\mu$.
For this reason, when correlating two differently biased samples of galaxies, the correlation $\langle \Delta_N \Delta_R \rangle$ sources odd multipoles of the power spectrum or 2-point correlation function.

We want to emphasise that this approach is purely geometrical and it can be applied to any metric theory of gravity. It therefore provides an appropriate framework to test gravity and the Equivalence Principle~\cite{Bonvin:2018ckp} at the largest scales. Furthermore, we only work within the weak field approximation neglecting terms directly proportional to the metric perturbations. Any contributions to the odd power spectrum multipoles beyond the weak field approximation are suppressed by $\epsilon_\HH^{3/2}$.

%%%%%%%%%%%
\section{Redshift perturbation}
\label{sec:redshift}

The first step in our approach consists in determining the redshift perturbation within the weak field approximation. We remind the reader that we do not intend to expand perturbatively around the matter density fluctuation $\delta_m$, which can be order $1$ in our derivation. We instead assume that the metric perturbation are small at any scale, i.e.~$\Psi \sim \Phi \ll 1$.
We start considering a perturbed FLRW metric
\be \label{eq:metric}
ds^2 = a\left( t \right)^2 \left[ -\left( 1+2 \Psi \right) dt^2 + \left( 1 - 2 \Phi \right) \left( dr^2 + r^2 d \Omega \right) \right] \, ,
\ee
where $\Psi$ and $\Phi$ denote (to first order) the Bardeen potentials and $t$ is the conformal time. 
In an arbitrary space-time the observed redshift is defined as
\be
1+z = \frac{k^\mu u_\mu  |_s}{k^\mu u_\mu  |_o}\, ,
\ee
where the suffixes $s$ and $o$ denote the source and the observer positions, respectively.
The 4-vector $k^\mu = dx^\mu /d\lambda$ is the tangent vector along the light geodesic connecting the source and the observer, determined by the geodesic equations
\be \label{geod_eq}
\frac{d k^\mu}{d\lambda} + \Gamma^\mu_{\nu \lambda} k ^\nu k^\lambda  = 0 \, , 
\ee
where $ \Gamma^\mu_{\nu \lambda}$ are the Christoffel symbols associated to the metric~\eqref{eq:metric}.
The 4-velocity $u^\mu$ is time-like and normalized such that
\be \label{eq:4vel_norm}
u^\mu u _\mu = -1 \, .
\ee
By writing the 4-velocity as $\left( u^\mu \right) =a^{-1} \left( u^0, v^i \right)$, where $\bv$ is the 3-dimensional peculiar velocity, we can fully constrain the $u^0$ component from eq.~\eqref{eq:4vel_norm}
\be \label{eq:u0}
u^0 = \sqrt{\frac{1+ \left(  1- 2 \Phi \right) v^2}{1+ 2 \Psi}} = 1 + \frac{v^2}{2} - \Psi + \mathcal{O} \left( \epsilon_\HH^2 \right) \, ,
\ee
where we have expanded in the weak field parameter\footnote{For sake of simplicity we always denote the weak field parameter $\epsilon_\HH$ in Fourier space. As specified before, in real space the weak field parameter $\epsilon_\HH$ is related to the inverse of the Laplacian operator. } $ \epsilon_\HH$. Within this scheme\footnote{This scheme has been theoretical motivated in Refs.~\cite{Green:2010qy,Green:2011wc}, and later adopted in several other works, see for instance Refs.~\cite{Adamek:2013wja,Adamek:2014gva,Adamek:2016zes,Fidler:2017pnb,Castiblanco:2018qsd}.} we have $\Psi \sim \Phi \sim v^2 \sim \epsilon_\HH \ll 1$. This approximation holds even after shell crossing in structure formation. Indeed the collapse of structures is counteracted by the virialized motion, which predicts $\Psi \sim \Phi \sim v^2$. While the amplitude of over-densities becomes hundred times the mean density of the universe in virialized object, the gravitational potentials are small everywhere and they become order unity only in the proximity of black holes.

By using the solution to the geodesic eq.~\eqref{geod_eq}
\be
\delta k^0 - \delta k^0_o =2 \Psi_o - 2 \Psi +\mathcal{O} \left( \epsilon_\HH^2 \right)
\ee
where we have neglected the integrated Sachs-Wolfe (ISW) contribution and the suffix $o$ denotes the quantities evaluated at the observer position, we obtain the observed redshift
\be \label{eq:redshift1}
1+ z = \left( 1 + \bar z \right) \left( 1 +  \left[- \ndv - \Psi+ \frac{v^2}{2}   \right]^{ \bar z}_o \right)  + \mathcal{O} \left( \epsilon_\HH^{3/2} \right) 
\ee
where we have introduced the background redshift $\bar z$ through $1+ \bar z = 1/a(t)$.

The redshift perturbation is defined as 
\be
\delta z = z - \bar z.
\ee
By plug in the redshift perturbation $\delta z$ and Taylor expanding around the observed redshift $z$, under the assumption $\delta z \ll 1$ we obtain
\bea
\delta z &=&  \left( 1+ \bar z\right) \left[ \frac{v^2}{2} -\ndv - \Psi \right]^{\bar z =z- \delta z}_o  + \mathcal{O} \left( \epsilon_\HH^{3/2} \right)
\nonumber \\
&=&
\sum_{i=0} \frac{1}{i!}   \left( - \delta z \right)^i    \frac{d^i }{dz^i} \left[ \left( 1+ z\right)  \left(  \frac{v^2}{2} - \ndv- \Psi - \frac{v_o^2}{2} + {\ndv}_o+ \Psi_o\right) \right] + \mathcal{O} \left( \epsilon_\HH^{3/2} \right) \, .
\label{eq:redshift3}
\eea
where we have considered\footnote{Since $\frac{d A\left( t , r \right)}{dz} = \frac{dt}{dz} \partial_t A +\frac{dr}{dz} \partial_r A  $, in the redshift derivative there is also a contribution of the order $\epsilon_\HH^0$, however this is subdominant with respect to the radial derivative which is of the order $\epsilon_\HH^{-1/2}$. Therefore considering $\frac{d^i}{dz^i} \sim \epsilon_\HH^{-i/2}$ is correct up to subleading terms in our perturbative expansion.} $\frac{d^i}{dz^i} \sim \epsilon_\HH^{-i/2}$ and $\delta z \sim \epsilon_\HH^{1/2}$.
We now make the following Ansatz
\be
\delta z = \sum_i \left. \delta z \right|_i \qquad \text{where} \quad \left. \delta z \right|_i =  \mathcal{O} \left(  
\left[  \left( 1+ z\right)  \left(  \frac{v^2}{2} - \ndv- \Psi  - \frac{v_o^2}{2} + {\ndv}_o+ \Psi_o \right) \right]^i \right) \, .
\ee
With this Ansatz and eq.~\eqref{eq:redshift3} we can determine any term $\left. \delta z \right|_i$ as a function of all other contributions $\left. \delta z \right|_j$ with $j<i$. 
For instance we have
\bea
\left. \delta z \right|_1 &=&  \left( 1+ z\right)  \left(  \frac{v^2}{2} - \ndv- \Psi  - \frac{v_o^2}{2} + {\ndv}_o+ \Psi_o \right) \, ,\\
%%%%%%%%%%%%%%%%%%%%%%%
\left. \delta z \right|_2 &=& -\left. \delta z \right|_1 \frac{d \left. \delta z \right|_1 }{dz} \, , \\
%%%%%%%%%%%%%%%%%%%%%%%
\left. \delta z \right|_3 &=& -\left. \delta z \right|_2 \frac{d \left. \delta z \right|_1 }{dz} +\frac{1}{2} \left( \left. \delta z \right|_1 \right)^2 \frac{d^2 \left. \delta z \right|_1 }{dz^2}\, , \\
%%%%%%%%%%%%%%%%%%%%%%%
\left. \delta z \right|_4 &=& -\left. \delta z \right|_3 \frac{d \left. \delta z \right|_1}{dz} + \left. \delta z \right|_1 \left. \delta z \right|_2 \frac{d^2\left. \delta z \right|_1}{dz^2}-\frac{1}{6}\left( \left. \delta z \right|_1\right)^3 \frac{d^3 \left. \delta z \right|_1}{dz^3} \, .
\eea
Moreover, we can re-sum all the contributions  $\left. \delta z \right|_i$, such that we can write the full redshift perturbation as
 \be
 \label{eq:DELTAZ}
\delta z = - \sum_{i=0} \frac{1}{\left( i+1\right)!} \frac{d^i}{dz^i} \left[ \left( 1+z \right) \left( \Psi + \ndv - \frac{v^2}{2} - \Psi_o- {\ndv}_o+ \frac{v_o^2}{2}  \right) \right]^{i+1}  + \mathcal{O} \left( \epsilon_\HH^{3/2}  \right)\, ,
\ee
which satisfies eq.~\eqref{eq:redshift3}.
The three contributions to the redshift-space distortion are due to gravitational redshift as well as the linear and transverse Doppler effect. Clearly, beyond linear theory these three terms are coupled together.

%%%%%%%%%%%%%%%%
\section{Density perturbation} 
\label{sec:density}

Now that we have derived the redshift perturbation $\delta z$, we want to study its relation to the observed galaxy density.
In this section we are therefore interested in the relativistic corrections to the density part of the galaxy number counts. We start considering the observed density $\rho_{\rm obs} \left( z , \bn \right)$ as a function of the measured redshift $z$ and photon direction $\bn$. This is related to the theoretical density $\rho$ as
\be \label{eq:den1}
\rho_{\rm obs} \left( z , \bn \right) = \rho_{\rm obs} \left( \bar z + \delta z , \bn^{(0) }+ \delta \bn \right) = \rho \left( \bar z, \bn^{(0)} \right).
\ee
Since the perturbations of the photon direction are due to the deflection angle sourced by the lensing potential, we neglect its contribution along this derivation as motivated in the previous sections. 
A non-perturbative study of the impact of the deflection angle to the matter power spectrum~\cite{DiDio:2016kyh} and on the correlation function~\cite{Dodelson:2008qc} have been already performed.

In order to express the density perturbation as a function of the observed redshift $z$ we simply need to Taylor expand eq.~\eqref{eq:den1} 
\be \label{eq:deltaz_Taylor}
\delta_z = \sum_i  \frac{1}{\bar \rho} \frac{1}{i! } \frac{d^i}{dz^i} \left[ \bar \rho \left( 1+ \delta \right) \right] \left( - \delta z \right)^i
 - 1,
\ee
where we have rewritten $\rho = \bar \rho \left( 1 + \delta \right)$.   
Being the redshift perturbation $\delta z$ of the order $\epsilon^{1/2}$ we can approximate $\delta_z$ through 
\bea  \label{eq:deltaz_Taylor2}
\delta_z &=& 
 \sum_{i=0} \frac{1}{i!} \left(  \frac{-\delta z }{\HH \left( 1 + z \right) } \right)^i \left[ \partial_r^i  +  \HH i \left( \frac{  1-i}{2} \left( 1 - \frac{\dot\HH}{\HH^2} \right) +3 {- {b_e}}\right)\partial_r^{i-1}  - i \partial_r^{i-1} \partial_t  \right] \delta 
\nonumber \\
&&+
\mathcal{O} \left( \epsilon_\HH \right) \, ,
\eea
where we have introduced the evolution bias (see e.g.~\cite{Challinor:2011bk,DiDio:2013bqa}) 
\be \label{fevo}
{b_e}= 3 - \left( 1+ z \right) \frac{d \ln \bar \rho}{dz},
\ee
to account for different galaxy evolution.

A non-vanishing evolution bias indicates that the number of sources is not conserved in a comoving volume.
Clearly for Dark Matter, $\rho_{DM} \propto a(t)^{-3}$, we have ${b_e} = 0$. 
However biased tracers may have a different background density evolution.
Interestingly, while ${b_e}$ is a function of redshift, its derivative does not contribute to the (leading) relativistic number counts. Indeed the term $b'_{\rm evo} \left( z \right) $ is associated with the second derivative of the background density. However, acting in eq.~\eqref{eq:deltaz_Taylor} with more than one redshift derivative on the background density $\bar \rho$, it reduces the number of spatial derivatives on the density perturbation $\delta$, suppressing these terms beyond the weak field approximation.

\section{Volume perturbation} 
\label{sec:volume}

We now consider the contribution due to the volume perturbation.
We follow the derivation of Ref.~\cite{Bonvin:2011bg}  by considering the following volume element around a source with 4-velocity given by~\eqref{eq:u0}
\be
dV = \sqrt{-g} \epsilon_{\mu \nu \alpha \beta} u^\mu dx^\nu dx^\alpha dx^\beta = v\left( z, \theta, \varphi \right) dz d\theta d\varphi \, ,
\ee
where $\epsilon_{\mu \nu \alpha \beta} $ is the Levi-Civita anti-symmetric tensor.
In line with the derivation of Ref.~\cite{DiDio:2018zmk} we have
\bea
v_{\rm obs} \left( z, \theta \varphi \right) &=& \sqrt{-g} \epsilon_{\mu \nu \alpha \beta} u^\mu \frac{\partial x^\nu}{\partial z} \frac{\partial x^\alpha}{\partial \theta} \frac{\partial x^\beta}{\partial \varphi}
\nonumber\\
&=&
\left[ a^3 r^2 \sin \theta \right] \left( \bar z \right) \left\{ \left( 1 + \frac{v^2}{2} - \Psi \right) \frac{ dr \left( \bar z \right) }{dz} + \ndv \left( \bar z \right) \frac{dt \left( \bar z \right)}{dz} \right\}+ \mathcal{O} \left( \epsilon_\HH \right) 
\nonumber \\
&=& \left[ a^3 r^2 \sin \theta \right] \left( \bar z \right) \frac{dr}{d\bar z} \left( 1 - \frac{d \delta z }{d z} \right)
 \left( 1 + \frac{v^2}{2} - \Psi  - \ndv \right)_{z=\bar z} + \mathcal{O} \left( \epsilon_\HH \right) 
 \nonumber \\
&=& \left[ \frac{a^4 r^2 \sin \theta }{\HH} \right] \left( \bar z \right)\left( 1 - \frac{d \delta z }{d z} \right)
 \left( 1 + \frac{\delta z}{1+z} { - {\ndv}_o} \right) + \mathcal{O} \left( \epsilon_\HH \right) 
  \nonumber \\
  &=&
  \bar v  \left( 1- \frac{1}{\bar v} \frac{d \bar v}{dz } \delta z \right)\left( 1 - \frac{d \delta z }{d z} \right)
 \left( 1 + \frac{\delta z}{1+z} { - {\ndv}_o} \right) + \mathcal{O} \left( \epsilon_\HH \right)\, ,
\eea
where we have introduced the background volume element $\bar v = r^2 a^4 \sin \theta /\HH$.
Consistent with the previous approximation we have neglected the contribution due to angular Jacobian
\be
\left| \frac{\partial \left( \theta_s , \varphi_s \right)}{\partial \left( \theta_o , \varphi_o \right)} \right| \, .
\ee
Indeed, this leads to lensing effects, which we are neglecting in this approach.

\section{Full relativistic number counts}
\label{sec:full_counts}

Once having derived the density and the volume contributions to the number counts, we can simply combine them to get the full relativistic number counts as function of the redshift perturbation $\delta z$
 \begin{align}
 \label{eq:DELTA}
\Delta \left( \bn , z \right) =&\; 
{ \left( 1+ \delta_z \right) \frac{v_{\rm obs}}{\bar v}  - 1 }= 
\nonumber \\
=& \; \sum_{i=0} \frac{1}{i!} \left(  \frac{-\delta z }{\HH \left( 1 + z \right) } \right)^i \bigg[ \partial_r^i  +  \HH i \left( \frac{ 1-i}{2} \left( 1 - \frac{\dot\HH}{\HH^2} \right) +3{ - {b_e}} \right)\partial_r^{i-1}  
\nonumber \\
&  \hspace{4cm}
- i \partial_r^{i-1} \partial_t  \bigg] \left( 1 + \delta \right)
\nonumber \\
& \times
  \left( 1- \frac{1}{\bar v} \frac{d \bar v}{dz } \delta z \right)\left( 1 - \frac{d \delta z }{d z} \right)
 \left( 1 + \frac{\delta z}{1+z}  {-  {\ndv}_o}\right) -1 + \mathcal{O} \left( \epsilon_\HH \right) \, .
  \end{align}
 We stress, once again, that this expression is valid at any order in perturbation theory. We clearly see that it depends completely on the redshift perturbation $\delta z$.

We can further simplify this expression by decomposing the redshift perturbation in a Newtonian and a relativistic part
\be
\delta z = \delta z_N + \delta z_R +  \mathcal{O} \left( \epsilon_\HH^{3/2} \right) , \qquad \text{where} \quad  \delta z_R  \sim \epsilon^{1/2} \delta z_N \,.
\ee
In this case, neglecting the terms evaluated at the observer position for sake of simplicity, we obtain
\bea
\label{eq:deltazN}
\delta z_N  &=& -  \left( 1 + z \right)\sum_{i=0} \frac{\HH^{-i}}{\left( 1 + i \right)!}  \partial_r^i \ndv^{i+1} \, ,\\
\delta z_R  &=& - \left( 1+z \right) \sum_{i=0} \frac{\HH^{1-i}} {\left( i+1 \right)!} \left[ 
\frac{i}{2} \left( i+3 + \left( i -1 \right) \frac{\dot \HH}{\HH^2} \right)  \partial_r^{i-1} \ndv^{i+1}
\right.
\nonumber \\
&&
\left.
\hspace{3cm}
- \frac{i \left( i+1 \right)}{\HH} \partial_r^{i-1} \left( \ndv ^i \dotndv \right)
+ \frac{i+1}{\HH}\partial_r^i \left( \ndv^i \left( \Psi - \frac{v^2}{2} \right)
 \right)\right] \, .
\eea
By adopting this redshift decomposition we get
\bea
\label{eq:general_numbercounts}
\Delta \left( \bn , z \right) &=& 
\left( 1 - \frac{d \delta z }{d z} \right) \left\{
  \sum_{i=0} \frac{1}{i!} \left(  \frac{-\delta z }{\HH \left( 1 + z \right) } \right)^i
 \partial_r^i \left( 1 + \delta \right) \right.
 \nonumber \\
 &&
 \left.
+  \sum_{i=1} \frac{1}{i!} \left(  \frac{-\delta z_N }{\HH \left( 1 + z \right) } \right)^i
  \bigg[   \HH i \left( \frac{ 1-i}{2} \left( 1 - \frac{\dot\HH}{\HH^2} \right) +3{ - {b_e}} \right)\partial_r^{i-1}  
- i \partial_r^{i-1} \partial_t  \bigg]  \delta \right\}
\nonumber \\
&& \times
  \left( 1- \frac{1}{\bar v} \frac{d \bar v}{dz } \delta z_N  + \frac{\delta z_N}{1+z}\right)
-1 + \mathcal{O} \left( \epsilon_\HH \right) \, .
\eea

In the next sections we will show briefly that we recover correctly the known results up to third order in perturbation theory with this simple approach. As shown in eq.~\eqref{eq:DELTA}, our new approach can easily obtain the (leading) relativistic perturbation to higher orders in perturbation theory without the need to re-derive them from scratch at any perturbative order.

%%%%%%
\subsection{Linear Perturbation Theory}

In this section we will show that our approach does recover the well-known number count at linear order.
Within linear theory we can neglect the transverse Doppler effect and therefore the redshift perturbation is simply given by\footnote{ The equations shown and derived in these sections are valid at the perturbative order considered. For sake of simplicity we do not explicitly denote the perturbative order in each equation. This helps also to avoid any possible confusion with the weak field approximation expressed in terms of power of $\epsilon_\HH$.}  
\bea
\delta z &=& - \sum_{i=0} \frac{1}{\left( i+1\right)!} \frac{d^i}{dz^i} \left[ \left( 1+z \right) \left( \Psi + \ndv - \frac{v^2}{2} -{\ndv}_o- \Psi_o + \frac{v_o^2}{2}   \right) \right]^{i+1}  + \mathcal{O} \left( \epsilon_\HH^{3/2} \right) 
\nonumber \\
&=& - \left( 1+ z \right) \left( \Psi + \ndv  - \Psi_o -{\ndv}_o\right) + \mathcal{O} \left( \epsilon_\HH^{3/2} \right)  \, ,
\eea
and in particular
\bea
\delta z_N &=&  - \left( 1+ z \right)   \left( \ndv - {\ndv}_o \right)  \, , \\
\delta z_R &=&  - \left( 1+ z \right)  \left(  \Psi - \Psi_o \right)  \, .
\eea
The number counts read as 
\bea
\Delta^{(1)}\left( \bn , z \right)\! &=&\! \left( 1 + \delta - \frac{\delta z_N}{ \left( 1 +z \right)}  \left( 3 - {b_e}\right) \right)\!  \left( 1 - \frac{1}{\bar v} \frac{ d \bar v}{dz} \delta z_N - \frac{ d \delta z}{dz} + \frac{\delta z_N}{1+z}  { - {\ndv}_o}\right)\! -1 \!+ \!\mathcal{O} \left( \epsilon_\HH \right) 
\nonumber \\
&=& \!\delta + \frac{\delta z_N}{1+z} \left( 1-3 +{b_e} - \frac{1+z}{\bar v}\frac{d \bar v}{d v }  \right)- \frac{d \delta z }{dz}  { - {\ndv}_o}
+ \mathcal{O} \left( \epsilon_\HH \right) 
\nonumber \\
&=& \!\delta \!-\! \left(  \ndv{ - {\ndv}_o}\right)\!  \left( \!2 +{b_e} - \frac{\dot \HH}{\HH^2} - \frac{2}{\HH r}  \!\right)+ \ndv + \left( 1+ z \right) \frac{d \left( \ndv + \Psi \right)}{dz}  { - {\ndv}_o}
+ \mathcal{O} \left( \epsilon_\HH \right) 
\nonumber \\
&=& \!\delta +\left(  \ndv { - {\ndv}_o}\right) \! \left(  \frac{\dot \HH}{\HH^2} + \frac{2}{\HH r}-1 -{b_e}  \! \right)\! +\! \HH^{-1} \left( \partial_r \ndv + \partial_r \Psi - \dotndv \right)  { - {\ndv}_o}
\!+\! \mathcal{O} \left( \epsilon_\HH \right) 
\nonumber \\
&&
\eea
where we have used 
\be
\frac{1+z}{\bar v}\frac{d \bar v}{d v }=  \frac{\dot \HH}{\HH^2} + \frac{2}{\HH r} - 4 \, .
\ee
Considering the Euler equation
\be
 \dotndv + \HH \ndv - \partial_r \Psi = 0 
\ee
we recover the well-known relativistic number counts to linear order, first derived in Refs.~\cite{Yoo:2009,Yoo:2010,Bonvin:2011bg,Challinor:2011bk},  namely
\be
\Delta^{(1)}\left( \bn , z \right) = \delta + \HH^{-1} \partial_r \ndv + \left(  \ndv { - {\ndv}_o}\right)  \left(  \frac{\dot \HH}{\HH^2} + \frac{2}{\HH r} -{b_e}   \right)
+ \mathcal{O} \left( \epsilon_\HH \right) \, .
\ee

In our derivation the density fluctuation $\delta$ is defined in Newtonian gauge. In most of other references it is written in terms of the density fluctuation in synchronous comoving gauge. Being the difference between the density fluctuation defined in the two gauges of the order $\mathcal{O}\left( \epsilon_\HH \right)$ we can simply exchange them in our expression. The synchronous gauge offers also a more natural framework to include the galaxy bias, see for instance~\cite{Baldauf:2011bh,Jeong:2011as}.

%%%%%%
\subsection{Second order Perturbation Theory}

At second order the redshift perturbation reads as\footnote{ For the rest of the paper we will neglect the terms evaluated at the observer position to simplify the comparison with previous results in the literature.}
\be
\delta z^{(2)} = \delta z^{(1 \rightarrow 2)}+ \left( 1+ z\right) \left( \frac{v^2}{2} - \ndv^2 - \HH^{-1}  \partial_r \left( \ndv \Psi \right)   - \HH^{-1} \ndv \partial_r \ndv + \HH^{-1} \ndv \dotndv \right)
+ \mathcal{O} \left( \epsilon_\HH^{3/2} \right),
\ee
where $ \delta z^{(1 \rightarrow 2)}$ denotes the terms which appear already at first order, simply evaluated at second order. Indeed at any order we will need to include the same operators which appear to lower orders, simply evaluated at higher order in perturbation theory, e.g.~$\delta^{(2)}, v^{(2)}, \Psi^{(2)}$.  
We can further decompose the second order redshift perturbation in
\bea
\delta z_N^{(2)} &=&  \delta z_N^{(1 \rightarrow 2)} - \left( 1+ z \right)   \HH^{-1} \ndv \partial_r \ndv \, ,
\\
\delta z_R^{(2)} &=&  \delta z_R^{(1 \rightarrow 2)}+  \left( 1+ z\right) \left( \frac{v^2}{2} - \ndv^2 - \HH^{-1}  \partial_r \left( \ndv \Psi  \right)  + \HH^{-1} \ndv \dotndv \right) \, .
\eea
By using eq.~\eqref{eq:general_numbercounts} we obtain directly
\bea
\Delta_N^{(2)} \left( \bn , z \right) &=& 
\delta_g^{(2)} + \HH^{-1} \partial_r \ndv^{(2)} 
+ \HH^{-1} \partial_r \left( \ndv  \delta_g \right)  + \HH^{-2}  \partial_r \left( \ndv \partial_r \ndv \right) \, ,
\\
\Delta_R^{(2)}\left( \bn , z \right) &=& 
  \left( -1 +\frac{\dot \HH}{\HH^2} + \frac{2}{\HH r} - {b_e}\right)\left(  \ndv^{(2)} +\ndv \delta  \right) 
  - \HH^{-1} \dotndv^{(2)} - 2 \HH^{-2} \partial_r \ndv \dotndv 
\nonumber \\
&&
+ \left( -2 +3 \frac{\dot \HH}{\HH^2} + \frac{4}{\HH r} - 2 {b_e} \right) \HH^{-1} \ndv \partial_r \ndv
- 2 \HH^{-2} \ndv \partial_r \dotndv
\nonumber \\
&&
 - \HH^{-1} \dotndv \delta - \HH^{-1} \ndv \dot \delta +   \HH^{-1}  v^a \partial_a \ndv  
 + \HH^{-2}\Psi\partial_r^2 \ndv  + \HH^{-1} \Psi \partial_r \delta
\nonumber \\
&&
{
 + \HH^{-1} \partial_r \Psi^{(2)}  + 2 \HH^{-2} \partial_r \ndv \partial_r \Psi+ \HH^{-1} \delta \partial_r \Psi + \HH^{-2} \ndv \partial_r^2 \Psi
} \,.
\label{eq:rel2}
\eea
This simple approach fully agrees with the derivations of Refs.~\cite{DiDio:2014lka,DiDio:2018zmk,Clarkson:2018dwn}.

%%%%%%
\subsection{Third order Perturbation Theory}

Similar to previous sections, we show here how to obtain straightforwardly third order perturbation theory in a relativistic framework. Again, the key quantity to derive is the redshift perturbation 
\bea
\delta z_N^{(3)} &=&  \delta z_N^{(2 \rightarrow 3)} - \frac{ 1+ z }{6 \HH^2} \partial^2_r \ndv^3 \, ,
\\
\delta z_R^{(3)} &=&  \delta z_R^{(2 \rightarrow 3)} + 
\left( 1+ z \right) \bigg[ \HH^{-2} \partial_r \left( \dotndv \ndv^2 \right)   
- 2 \HH^{-2} \partial_r^2 \left( \ndv^2 \Psi \right) 
+ 2 \HH^{-1} \partial_r \left( \ndv v^2 \right) 
\nonumber \\
 &&  \hspace{3cm}
- \left( \frac{5}{2} + \frac{\dot \HH}{2 \HH^2} \right) \HH^{-1} \partial_r \ndv \ndv^2 
\bigg] \, .
\eea
Once we have derived the redshift perturbation we can plug it in eq.~\eqref{eq:general_numbercounts} to obtain
\bea
\Delta_N^{(3)}\left( \bn , z \right) &=&  { \delta_g^{(3)} +\frac{{\partial_r \ndvthree}}{\HH}}
{ 
+\left[ \HH^{-1} \partial_r\left(  \ndv\delta_g  \right)\right]^{(3)}
+ \left[ \HH^{-2}  \partial_r \left( \ndv \partial_r \ndv \right) \right]^{(3)}
}
\nonumber \\
&&
 + { \frac{1}{6} \HH^{-3}\partial_r^3 \ndv^3
 +\frac{1}{2} \HH^{-2} \partial_r^2 \left( \delta_g \ndv^2 \right)
 } \, ,
\\
\Delta_R^{(3)}\left( \bn , z \right) &=& \left\{
 \left( -1 +\frac{\dot \HH}{\HH^2} + \frac{2}{\HH r} - {b_e}\right)\left(  \ndv^{(3)} + \left[ \ndv \delta \right]^{(3)} \right) 
  - \HH^{-1} \dotndv^{(3)} 
  \right.
\nonumber \\
&&
- 2 \HH^{-2} \left[  \partial_r \ndv \dotndv \right]^{(3)}
+ \left( -2 +3 \frac{\dot \HH}{\HH^2} + \frac{4}{\HH r} - 2 {b_e} \right) \left[  \HH^{-1} \ndv \partial_r \ndv \right]^{(3)}
\nonumber \\
&&
- 2 \HH^{-2} \left[ \ndv \partial_r \dotndv \right]^{(3)}
 - \HH^{-1} \left[ \dotndv \delta \right]^{(3)} - \HH^{-1} \left[\ndv \dot \delta\right]^{(3)} +   \HH^{-1}  \left[v^a \partial_a \ndv\right]^{(3)}
\nonumber \\
&&
 + \HH^{-2} \left[ \Psi\partial_r^2 \ndv \right]^{(3)}  
+ \HH^{-1} \left[ \Psi \partial_r \delta\right]^{(3)}
 + \HH^{-1} \partial_r \Psi^{(3)}  + 2 \HH^{-2} \left[ \partial_r \ndv \partial_r \Psi \right]^{(3)}
 \nonumber \\
&&
\left.
+ \HH^{-1} \left[ \delta \partial_r \Psi\right]^{(3)}
 + \HH^{-2} \left[ \ndv \partial_r^2 \Psi\right]^{(3)}
\right\} \nonumber
\\
&& +
\frac{1}{2 \HH^3} \partial^3_r \left( \ndv^2 \Psi \right)
 - \frac{1}{2 \HH^3} \partial_t \partial_r^2 \ndv^3
 + \frac{1}{\HH^2} \partial_r^2 \left( \Psi \ndv \delta \right)
 - \frac{1}{2\HH^2} \partial_r^2 \left( \ndv v^2 \right)
 \nonumber \\
 &&
 - \frac{1}{\HH^2} \partial_t \partial_r \left( \delta \ndv^2 \right)
 - \frac{1}{2 \HH} \partial_r \left( \delta v^2 \right)
  + \frac{1}{\HH^2} \partial_r \left( \partial_r \ndv \ndv^2 \right) \left( 3 \frac{\dot\HH}{\HH^2} + \frac{3}{\HH r} - \frac{3}{2} {b_e} \right) 
  \nonumber \\
 &&
 + \frac{1}{\HH} \partial_r \left( \delta \ndv^2 \right) \left( - \frac{1}{2} + \frac{3}{2} \frac{\dot\HH}{\HH^2} + \frac{2}{\HH r} - {b_e} \right) \, .
\label{eq:rel3}
\eea
The terms in the curly brackets are the operators already present in $\Delta_R^{(2)}\left( \bn , z \right)$ simply evaluated at third order.
This result fully agrees with Ref.~\cite{DiDio:2018zmk}.

\section{Single contributions to galaxy number counts}
\label{sec:single_contributions}

So far we have shown that our new formalism agrees with the current literature.
In this section we want to fully leverage this formalism by deriving the contributions induced by the different effects of the redshift perturbation on the galaxy number counts.
Being interested in testing this approach at the next-to-leading order we derive the effect on the galaxy number counts up to third order in perturbation theory.
The purpose of this section is also to derive the terms that allow us to compute (in the accompanying paper~\cite{Beutler:2020}) the dipole induced by these effects and then compare with the RayGalGroup simulation~\cite{Breton:2018wzk}.
In this regard, we do not compute the contribution of the single terms alone, but we will follow the same redshift perturbation definitions adopted in Ref.~\cite{Breton:2018wzk}.

\subsection{Gravitational potential}
We start by considering the redshift perturbation induced by the gravitational potential, i.e.~the so-called gravitational redshift
\be
\delta z_1 = - \left( 1+ z \right) \Psi \, .
\ee
Clearly in this case we do not have any Newtonian contribution to the redshift perturbation and therefore eq.~\eqref{eq:general_numbercounts} simply reduces as (considering only terms up to third order in perturbation theory)
\bea
\Delta_1 \left( \bn , z \right) &=& 
\left( 1 - \frac{d \delta z_1 }{d z} \right) \left\{
  \sum_{i=0} \frac{1}{i!} \left(  \frac{-\delta z_1 }{\HH \left( 1 + z \right) } \right)^i
 \partial_r^i \left( 1 + \delta \right) \right\}
-1 + \mathcal{O} \left( \epsilon_\HH \right) 
\nonumber \\
&\simeq&
\left( 1+\HH^{-1} \partial_r \Psi \right) 
\left( 1 + \delta  + \HH^{-1} \Psi \partial_r \delta \right) -1 + \mathcal{O} \left( \epsilon_\HH \right) 
\nonumber \\
&\simeq&
\delta +  \HH^{-1} \partial_r \Psi + \HH^{-1} \delta   \partial_r \Psi +  \HH^{-1} \Psi \partial_r \delta 
+ \mathcal{O} \left( \epsilon_\HH \right)
\, .
\eea
At the different orders in perturbation theory we have therefore
\bea
\Delta_1^{(1)}\left( \bn , z\right) &=& \delta_g + \HH^{-1} \partial_r \Psi
\, ,\\
\Delta_1^{(2)}\left( \bn , z\right) &=& \Delta_1^{(1\rightarrow 2)} +  \HH^{-1} \partial_r \left( \delta_g \Psi \right)
\, ,\\
\Delta_1^{(3)}\left( \bn , z\right) &=& \Delta_1^{(2\rightarrow 3)}
\, ,
\eea
where the perturbations on the right-hand side are evaluated at linear order if not specified otherwise.

%%%%%%%%%%%%%%%%%%%%%%%%%%%%%%%%%%%
\subsection{Linear Doppler effect}
We now include the linear Doppler effect in redshift perturbation on top of the gravitational potential computed in the previous section
\be
\delta z_2 = -\left(1+z \right) \left(\ndv + \Psi \right) 
\ee
Since now we also have a non-vanishing Newtonian redshift perturbation sourced by the linear Doppler effect the derivation is slightly more involving. By plugging the redshift perturbation $\delta z_2$ in eq.~\eqref{eq:general_numbercounts} we finally obtain
\bea
\Delta_2^{(1)} \left( \bn , z \right) &=& \delta_g + \HH^{-1} \partial_r \ndv 
+ \HH^{-1} \partial_r \Psi - \HH^{-1} \dotndv + \ndv \mathcal{R}
\, \\
\Delta_2^{(2)} \left( \bn , z \right) &=& \Delta_2^{( 1 \rightarrow 2)} 
+ \HH^{-1} \partial_r \left( \delta_g \left( \Psi + \ndv \right) \right) - \HH^{-1}\partial_t \left( \delta_g \ndv \right) 
+ \delta_g  \ndv  \mathcal{R}
\nonumber \\
&&
+ \HH^{-1} \ndv \partial_r \ndv \left(\mathcal{R} - 1\right)
\, , \\
\Delta_2^{(3)} \left( \bn , z \right) &=& \Delta_2^{( 2 \rightarrow 3)} + \HH^{-1} \delta_g \ndv \partial_r \ndv \left( \mathcal{R} -1 \right) 
+ \HH^{-2} \partial_r \left(\ndv \Psi \partial_r \delta_g \right) 
- \HH^{-2} \partial_r \left( \ndv \dot \delta_g \right) \ndv 
\nonumber \\
&&
- \HH^{-2} \partial_r \delta_g \dotndv \ndv
+\frac{ \HH^{-1}}{2} \ndv^2 \partial_r \delta_g  \left(b_{e } + 3 \mathcal{R} - \frac{2}{\HH r}  \right) \, ,
\eea
where we have introduced the parameter
\be
\mathcal{R} = -1 -{b_e} +\frac{\dot \HH}{\HH^2} + \frac{2}{\HH r}
\, .
\ee
Since the galaxy number counts do not depend linearly on the redshift perturbation we obtain a coupling between the Doppler $\ndv$ and the gravitational potential $\Psi$ at third order in perturbation theory.

%%%%%%%%%%%%%%%%%%%%%%%%%%%%%%%%%%%
\subsection{Transverse Doppler effect}
We now include also the last effect which sources the redshift perturbation, namely the transverse Doppler term
\be
\delta z_3 =  -\left(1+z \right) \left(\ndv + \Psi - \frac{v^2}{2}\right) \,.
\ee
Similarly to the previous section we plug the redshift perturbation $\delta z_3$ into eq.~\eqref{eq:general_numbercounts} obtaining 
\bea
\Delta_3^{(1)} \left( \bn , z \right) &=& \Delta_2^{(1)} 
\, , \\
\Delta_3^{(1)} \left( \bn , z \right) &=& \Delta_2^{(2)}  - \HH^{-1} \bv \cdot \partial_r \bv
\, , \\
\Delta_3^{(1)} \left( \bn , z \right) &=& \Delta_2^{(3)}  - \HH^{-1} \left[  \bv \cdot \partial_r \bv\right]^{(3)} - \frac{\HH^{-1}}{2} \partial_r \left( \delta_g v^2 \right) 
\, .
\eea
Since the transverse Doppler effect is a second order effect, the linear galaxy number counts fully agree with the one derived in the previous section for $\delta z_2$.

%%%%%%%%%%%%%%%%%%%%%%%%
\section{Newtonian limit}
\label{sec:newt_limit}

In this section we show that the formalism, we developed in this manuscript, recovers non-perturbatively the correct Newtonian limit. In previous sections we have already shown this up to third order in perturbation theory.
In the Newtonian limit we have
\be \label{eq:newt_limit1}
\Delta_N \left( \bn , z \right) = 
 \left(  \sum_{i=0} \frac{1}{i!} \left(  \frac{-\delta z_N }{\HH \left( 1 + z \right) } \right)^i  \partial_r^i   \left( 1+ \delta \right) 
\right)
\left( 1 - \frac{d \delta z_N }{d z} \right) - 1.
\ee
We aim to show that this equation agrees with (see Ref.~\cite{Vlah:2012ni}) 
\be \label{eq:deltaS_Vlah}
\delta_s \left( \bx \right) =  \sum_{L=0} \frac{1}{L!} \HH^{-L} \partial_r^L \left[ \left( 1 + \delta \left( \bx \right) \right) \ndv \left( \bx \right)^L \right] \, .
\ee
We can re-express this result in terms of sum over different perturbative order $n$
\be \label{eq:newt_limit2}
\delta_s \left( \bx \right) =
  \sum_{n=0} \frac{1}{n!} \HH^{-n} \partial_r^n \ndv^n
  +
  \sum_{n=0} \frac{1}{\left( n-1 \right)!} \HH^{-n+1} \partial_r^{n-1} \left( \delta  \ndv^{n-1} \right)
 \, .
\ee
Then rewriting eq.~\eqref{eq:newt_limit1} as 

\bea \label{eq:7.4}
\Delta_N \left( \bn , z \right) &=& 
 \left( 1+ \sum_{i=0} \frac{1}{i!} \left(  \frac{-\delta z_N }{\HH \left( 1 + z \right) } \right)^i  \partial_r^i   \delta  
\right)
\left( 1 - \frac{d \delta z_N }{d z} \right) - 1
=
\nonumber \\
&=& - \frac{d \delta z_N }{d z} + \sum_{i=0} \frac{1}{i!} \left(  \frac{-\delta z_N }{\HH \left( 1 + z \right) } \right)^i  \partial_r^i   \delta  
\left( 1 - \frac{d \delta z_N }{d z} \right) = 
\nonumber \\
&=& \sum_{i=0} \frac{\HH^{-i} }{i!} \partial_r^i \ndv^i - 1 
+ \sum_{i=0} \frac{1}{i!} \left(  \frac{-\delta z_N }{\HH \left( 1 + z \right) } \right)^i  \partial_r^i   \delta \left( 1 - \frac{d \delta z_N }{d z} \right)
\eea
we remark that we just need to show 
\be \label{eq:toshow}
  \sum_{n=0} \frac{ \HH^{-n+1}}{\left( n-1 \right)!} \partial_r^{n-1} \left( \delta  \ndv^{n-1} \right) =
   \sum_{i=0} \frac{1}{i!} \left(  \frac{-\delta z_N }{\HH \left( 1 + z \right) } \right)^i  \partial_r^i   \delta \left( 1 - \frac{d \delta z_N }{d z} \right) -1\, .
\ee
By plugging in the redshift perturbation~\eqref{eq:deltazN}, the right-hand side reads as
\bea
  && \sum_{i=0} \frac{1}{i!} \left(  \frac{-\delta z_N }{\HH \left( 1 + z \right) } \right)^i  \partial_r^i   \delta \left( 1 - \frac{d \delta z_N }{d z} \right) -1
=
\nonumber \\
&&=\sum_{i=0} 
\frac{1}{i!} \left( \sum_{\ell=0} \frac{\HH^{-\ell-1}}{\left( 1 + \ell \right)!}  \partial_r^\ell \ndv^{\ell+1}  \right)^i   \partial_r^i   \delta
\left( \sum_{j=0} \frac{\HH^{-j}}{j!} \partial^j \ndv^j \right) - 1 
\nonumber \\
&&=\sum_{i=0, j=0}  \frac{\HH^{-j-i} }{i! j!}   \partial_r^i   \delta \partial^j \ndv^j \left( \sum_{\ell=0} \frac{\HH^{-\ell}}{\left( 1 + \ell \right)!}  \partial_r^\ell \ndv^{\ell+1}  \right)^i -1.
\eea
We can also rewrite it as a sum over all perturbative orders $n$, like the left-hand side of eq.~\eqref{eq:toshow}.
 In this case, at a given perturbative order $n$, the internal sum over the index $\ell$ is bounded by $n -1-j$, and therefore we 
have
\be \label{eq:7.7}
\left[ 
\sum_{i=0, j=0}  \frac{\HH^{-j-i} }{i! j!}   \partial_r^i   \delta \partial^j \ndv^j \left( \sum_{\ell=0}^{n -1-j} \frac{\HH^{-\ell}}{\left( 1 + \ell \right)!}  \partial_r^\ell \ndv^{\ell+1}  \right)^i
\right]^{(n)} = 
\sum_{i=0}^{n-1} \sum_{j=0}^{n-i-1}  \frac{\HH^{-j-i} }{i! j!}   \partial_r^i   \delta \partial^j \ndv^j  G^{({ n-1-j})}_{ij},
\ee
where we have used the multinomial theorem obtaining 
\bea
G^{({ n-1-j})}_{ij} &\equiv& \left[ \left( \sum_{\ell=0}^{n-1-j} \frac{\HH^{-\ell}}{\left( 1 + \ell \right)!}  \partial_r^\ell \ndv^{\ell+1}  \right)^i \right]^{({ n-1-j})} \!\!\!\!\!\!\!\!\!=\!\!\!\!\!\!\!\!\!\!\!\!\!\!\! 
\sum_{\substack{k_0+..+k_{n-1}=i \\ \sum_{\ell=0}^{\ell-1} \left( \ell +1 \right) k_\ell = n-1-j} }\!\!\!\!\!\!
\frac{i!}{k_0! .. k_{n-1}!} \prod_{t=0}^{n-1} \left( \frac{\HH^{-t}}{\left( t+1 \right)!} \partial_r^t \ndv^{t+1} \right)^{k_t}
\nonumber \\
&=&
\HH^{1+j+i-n} 
\left( \begin{array}{c}
n-j-2 \\ i-1 \end{array} \right) \frac{i!}{\left( n-j -1 \right)!}
\partial_r^{n-1-j-i} \ndv^{n-1-j} 
\nonumber \\
&=&
 \frac{i}{n-j -1 } \frac{\HH^{1+j+i-n} }{\left( n-j-i -1 \right)!}
\partial_r^{n-1-j-i} \ndv^{n-1-j} \, .
\eea
So combining with eq.~\eqref{eq:7.7} we simply get
\bea
&& \hspace{-0.5cm}
\left[ 
\sum_{i=0, j=0}  \frac{\HH^{-j-i} }{i! j!}   \partial_r^i   \delta \partial^j \ndv^j \left( \sum_{\ell=0}^{n -1-j} \frac{\HH^{-\ell}}{\left( 1 + \ell \right)!}  \partial_r^\ell \ndv^{\ell+1}  \right)^i
\right]^{(n)} 
= 
\nonumber \\
&&=
\sum_{i=0}^{n-1} \sum_{j=0}^{n-i-1} \frac{\HH^{1-n}}{\left( i -1 \right)! j! } \frac{1}{\left( n-j -1 \right) \left( n-j-i -1 \right)!} \partial_r^i \delta \partial_r^j \ndv^j \partial_r^{n-1-j-i} \ndv^{n-1-j}.
\eea
We expand now the left-hand side of eq.~\eqref{eq:toshow} 
\bea
\frac{ \HH^{-n+1}}{\left( n-1 \right)!} \partial_r^{n-1} \left( \delta  \ndv^{n-1} \right)  &=& \frac{ \HH^{-n+1}}{\left( n-1 \right)!} \sum_{i=0}^{n-1} 
\left( \begin{array}{c} n-1 \\ i  \end{array} \right) \partial_r^i \delta \partial_r^{n-1-i} \ndv^{n-1}
\nonumber \\
&=& \sum_{i=0}^{n-1} 
 \frac{ \HH^{1-n}}{i! \left( n-1 -i \right)!}  \partial_r^i \delta \partial_r^{n-1-i} \ndv^{n-1}.
\eea
To conclude this proof we use
\be
\partial^M_r \ndv^{N} = \sum_{j=0}^M 
\frac{M-N}{j -N} 
\left( \begin{array}{c}
M \\ j
\end{array}\right)
 \partial_r^j \ndv^j \partial_r^{M-j} \ndv^{N-j}
\ee
for $N\geq 1$ and $M\leq N-1$. Indeed, we simply need to replace $M= n-i -1$ and $N=n-1$ to proof the validity of eq.~\eqref{eq:toshow}. Therefore, we have formally shown that this approach recovers the standard Newtonian perturbation theory in redshift space to any order in perturbation theory.

%%%%%%%%%
\section{Conclusions}
\label{sec:conclusions}

In this paper we have derived a novel approach to compute the (leading) relativistic corrections to the galaxy number counts to any perturbative order. We fully describe the relativistic effects, see eq.~\eqref{eq:DELTA}, in terms of redshift perturbation, eq.~\eqref{eq:DELTAZ}, offering a neat interpretation of the output of relativistic N-body simulations (through geodesic light-tracing~\cite{Breton:2018wzk}). 
As in N-body simulations, where the angular and redshift positions need to be corrected by the observed incoming photon direction and redshift perturbation by solving the geodesic propagation equations, we can write the relativistic effects in terms of the deflection angle and redshift perturbation. In our work we have focused on the redshift perturbation since it is the main source of the odd multipoles of the power spectrum or correlation function, while the deflection angle is sourced by the well-known lensing effect.

We have shown that this approach agrees up to third order\footnote{Previous derivations have stopped to third order.} with previous derivations of relativistic effects. 
We, for the first time, quantified the impact of the gravitational potential as well as linear and transverse Doppler effects on the galaxy number counts.
Moreover, we have proven that the Newtonian contributions are equivalent to standard perturbation theory to any order.

In an accompanying paper~\cite{Beutler:2020}, we have applied and tested this formalism with relativistic N-body simulations provided by~\cite{Breton:2018wzk}.

\acknowledgments
ED (No.~171494 and~171506) acknowledges financial support from the Swiss National Science Foundation.
FB is a Royal Society University Research Fellow.

\bibliographystyle{JHEP}
\bibliography{biblio_dipole}

\end{document}